\documentstyle[12pt]{article}
 
\def\a{\alpha} 
\def\b{\beta} 
\def\d{\delta} 
\def\e{\epsilon}

\def\j{\psi}
\def\k{\kappa} 
\def\l{\lambda} 
\def\m{\mu} 
 
\def\o{\omega} 
 
\def\q{\theta} 

\def\s{\sigma} 
\def\t{\tau}

\def\z{\zeta} 
 
\def\F{\Phi} 
\def\G{\Gamma}
 
\def\Ld{\Lambda} 
\def\O{\Omega} 

\def\X{\Xi} 

\def\cv{{\cal V}}

\def\inbar{\vrule height1.5ex width.4ptdepth0pt} 
\def\rlx{\relax\leavevmode} 
\def\I{\leavevmode
\hbox{\small1\kern-3.8pt\normalsize1}}
\def\openone{\leavevmode\hbox{\small1\kern-3.3pt\normalsize1}} 
\def\Ione{\rlx{\rm 1\kern-2.7pt l}}
\font\cmss=cmss10 
\font\cmsss=cmss10 at 7pt 
\def\ZZ{\rlx\leavevmode \ifmmode\mathchoice {\hbox{\cmss
Z\kern-.4em Z}} {\hbox{\cmss Z\kern-.4em Z}} 
{\lower.9pt\hbox{\cmsss Z\kern-.36em Z}} 
{\lower1.2pt\hbox{\cmsssZ\kern-.36em Z}} 
\else{\cmss Z\kern-.4em Z}\fi} 
\def\Ik{\rlx{\rm I\kern-.18em k}} 
\def\IC{\rlx\leavevmode 
\ifmmode\mathchoice {\hbox{\kern.33em\inbar\kern-.3em{\rm C}}}
{\hbox{\kern.33em\inbar\kern-.3em{\rm C}}} 
{\hbox{\kern.28em\sinbar\kern-.25em{\rm C}}}
{\hbox{\kern.25em\ssinbar\kern-.22em{\rm C}}} 
\else{\hbox{\kern.3em\inbar\kern-.3em{\rm C}}}\fi} 
\def\IP{\rlx{\rmI\kern-.18em P}} 
\def\IR{\rlx{\rm I\kern-.18em R}} 
\def\IN{\rlx{\rm I\kern-.20em N}}

\newcommand{\ol}\overline 
\newcommand{\ti}\tilde 
\newcommand{\wt}\widetilde 
\newcommand{\wh}\widehat
\newcommand{\bv}\breve 
\newcommand{\dg}\dagger

\newcommand{\aand}{\;\;\;\mbox{and}\;\;\;}

\newcommand{\be}{\begin{equation}} 
\newcommand{\ee}{\end{equation}}
\newcommand{\bl}{\begin{eqnarray}&} 
\newcommand{\el}{&\end{eqnarray}} 
\newcommand{\bq}{\begin{eqnarray}}
\newcommand{\eq}{\end{eqnarray}} 
\newcommand{\0}{{\bf 0}}

\newcommand{\ad}{{\dot\alpha}} 
\newcommand{\bd}{{\dot\beta}}

\newcommand{\uptad}{\widetilde\theta^{\dot\alpha}}

\newcommand{\qwt}{\widetilde\theta} 
\newcommand{\ov}{\overline}
\newcommand{\pa}{\partial}

\def\sl#1{\rlap{\hbox{$\mskip 1 mu /$}}#1} 

\begin{document} 
\title{\bf The Geometry of N=1 and N=2 Real Supersymmetric 
Non-Linear $\s$-Models \\in the
Atiyah-Ward Space-Time} 
\author{{\it M. Carvalho} 
{\thanks{e-mail: 698m5051@mn.waseda.ac.jp}}, 
\\ Waseda University 
\\ Department of Mathematics \\ 3-4-1
Okubo, Shinjuku-ku, Tokyo 169, Japan \\ 
\\ {\it J.A. Helay\"el-Neto} {\thanks{e-mail: helayel@cbpfsu1.cat.cbpf.br}}, 
{\it M.W. de Oliveira} {\thanks{e-mail: mwerneck@cbpfsu1.cat.cbpf.br}} 
\\ Centro Brasileiro de Pesquisas F\'\i sicas (CBPF)
\\ Departamento de Teoria de Campos e Part\'\i culas (DCP) 
\\ Rua Dr. Xavier Sigaud, 150 - Urca \\ 
22290-180 Rio de Janeiro - RJ - Brazil.} 
\date{}
 \maketitle 
 \begin{abstract} 
 {We analyse the structure of N=1 and N=2
supersymmetric non-linear $\s$-models built 
up with a pair of real superfields defined 
in the superspace of
Atiyah-Ward space-time. The geometry arising 
has new features such as the existence of a 
locally product structure
(N=1 case) and a set of automorphisms of the 
tangent space that is isomorphic to the 
split-quaternionic algebra (N=2 case).} 
\end{abstract} 
\newpage 
 \section{Introduction}
Bosonic non-linear (N=0) $\s$-models come in
evidence in Physics as the structure of scalar 
fields that appear in a theory with a 
spontaneously broken symmetry.
In all these models, the scalars define a 
mapping from the underlying space-time into 
a Riemannian manifold parametrised by coordinates 
that are the scalar fields themselves. Non-linear 
supersymmetric $\s$-models are then
natural generalisations of the bosonic ones in 
which the scalar fields are now N=1 superfields, 
i.e., they provide a representation of N=1 supersymmetry. 
The first formulation of supersymmetric $\s$-models in 
superspace was given by Zumino {\cite{zumino}}. 
He succeded to write a supersymmetric invariant 
action for a N=1, D=(3+1) model. In that reference, 
it was shown that the scalar superfields 
span a K\"ahler manifold. Later on, Hull et all 
{\cite{hull}} also succeded to write
a N=2 supersymmetric $\s$-model in superspace. 
They showed how to derive a N=2 supersymmetric
action over a Hyperk\"ahler manifold making use of its 
quaternionic structures.{\footnote{In ref.{\cite{gaume}} the
same ideas were developed working with the component 
approach and in 2 dimensions.} The same constructions can
also be performed for a space-time with signature D=(2+2), 
the so-called Atiyah-Ward space-time
\cite{marcelo1},\cite{marcelo2}. The common aspect in all 
these constructions is the use of a pair $(\phi,\phi^{*})$
of complex scalar superfields and their conjugates. 
The fact that the double 
covering of the isometry group of Atiyah-Ward space-time
is $SL(2,R)$ bring us new features to the formulation of 
supersymmetric models \cite{gates1}, \cite{oswaldo}. Here the
supersymmetric chiral and antichiral sectors, for example, 
are no more related by complex conjugation. Hence, we can work
work consistently with a pair of real 
scalar superfields of different chiralities. We can also define
Majorana-Weil spinors. We named such models as real 
supersymmetric $\s$-models. As a result, we obtain a geometry
that is different from that one which appears in
the formulation with a pair of complex superfields and their 
conjugates: in the N=1 case we obtain
a manifold that presents some characteristics between a 
K\"ahler and a locally-product manifold (the latter have already
been found by Gates et all in their formulation of 
twisted supersymmetric $\s$-models \cite{gates2}), 
while in the N=2 case we obtain a manifold that admits a set of 
automorphisms of the tangent bundle that is parametrized by the
split-quaternionic algebra. This paper is organised 
as follows: in Sections 2 and 3, we shall show how physical
requirements determine the characteristics of the 
geometry of the N=1 and N=2 supersymmetric $\s$-models,
respectively. In Section 4, we review some definitions 
concerning the split-quaternionic algebra, we establish 
definitions similar to the quaternion case and finally 
we present a mathematical formulation of our results.
\section{The Real N=1 Supersymmetric $\s$-Model}
Let us consider a set of 2n real superfields
$X^{I}\equiv(X^{i},X^{\hat i})=(\F^{i},\X^{i}), 
{\hat i}=i+n ;\;\;i=1,...,n,$ respectively chiral and 
antichiral whose expansion in
components read off as (we are using the same notations 
of {\cite{oswaldo}}), 
\be 
\F^{i}= A^{i} + 
i\q\j^{i} + i\q^{2} F^{i}
+ i\tilde\q \tilde{\sl{\pa}}\q A^{i} +
\frac{1}{2}\q^{2}\tilde\q \tilde{\sl{\pa}}\j^{i}
-\frac{1}{4}\q^{2}\tilde\q^{2}{\partial_{\mu } 
\partial^{\mu }} A^{i} \;\; , 
\ee 
\be 
\X^{i}= B^{i} + 
i\tilde\q\tilde\chi^{i} +
i\tilde\q^{2} G^{i} + i\q \sl{\pa}\tilde\q B^{i}+
\frac{1}{2}\tilde\q^{2}\q \sl{\pa} \tilde \chi^{i} -
\frac{1}{4}\q^{2}\tilde\q^{2}
{\partial_{\mu } \partial^{\mu }} B^{i}\;\;\;, 
\ee 
where $A^{i}$ and $B^{i}$ are real scalar fields, 
$\psi^{i}$ and
${\tilde{\chi}}^{i}$ are Majorana-Weyl spinors and 
$F^{i}$ and $G^{i}$ are real scalar auxiliary fields. 
A scalar superfield
is chiral ($\F$) or antichiral ($\X$) if it 
 satisfies respectively 
\bq 
&{\wt{D}_{\ad}}\F^{i}=0\;,\;\;\;
{\wt{D}_{\ad}}\F^{*i}=0\;,\;\;\aand \;\; 
D_{\a}\X^{i}=0\;,\;\;\; 
D_{\a}\X^{*i}=0 \;, 
\label{quiral} 
\eq 
with 
\bq 
&D_{\a}=\pa_{\a}-i{\pa}_{\a \ad}\uptad\;,\;\;\; 
{\wt{D}_{\ad}}=\wt{\pa}_{\ad}-i{\wt{\pa}}_{\ad \a} \q^{\a}\;, \\
&\{D_{\a},{\wt{D}_{\ad}}\}=-2i\;{\s}^{\m}_{\a \ad}\;
{\pa}_{\m}\;,\;\{D_{\a},{{D_{\b}}}\}=\{{\wt{D}_{\ad}}, 
{\wt{D}_{\bd}}\}=0\;\;, 
\nonumber 
\\ &[D_{\a},{\pa}_{\m}]=[{\wt{D}_{\ad}},{\pa}_{\m}]=0\;\; , 
\label{dalgebra} 
\nonumber 
\eq 
Following Zumino's work \cite{zumino} 
we write the action for the non-linear 
$\s$-model as, 
{\footnote{$\int{d^4xd^2{\q}d^2\qwt}
\equiv \frac{1}{16}\int{d^4x}D^{\a} 
{\wt{D}^{\ad}} \wt{D}_{\ad}D_{\a}$}} 
\be
S=2\int{d^4x~d^2{\q}~d^2\qwt}~K(\F^{i},\X^{i}) 
\label{action}\;, 
\ee 
$K$ being a real scalar function of the 2n
chiral/antichiral superfields. After eliminating 
the auxiliary fields $F^{i}, G^{i}$, by using their 
equations of motion, we
have the action expressed in component form as 
\be 
S=\int{d^4x~d^2{\q}~d^2\qwt}\{ 2g_{i{\hat j}}\pa_{\mu}A^{i}
\pa^{\mu}B^{j} -\frac{i}{2} g_{i{\hat j}} 
\psi^{i}\s^{\mu}D_{\mu}{\tilde \chi}^{j} -
\frac{i}{2}g_{i{\hat j}}{\tilde \chi}^{j}{\tilde \s}^{\mu}D_{\mu}\psi^{i} + 
\frac{1}{8}R_{i{\hat m}j{\hat n}}{\tilde \chi}^{m}{\tilde \chi}^{n} 
\psi^{i}\psi^{j}\} \;, 
\label{action1} 
\ee 
with
\bq
D_{\mu}\psi^{i}= \pa_{\mu}\psi^{i} + g^{i\hat k}\pa_{l}g_{r\hat k}
\psi^{r}\pa_{\mu}A^{l}\nonumber \\
D_{\mu}{\tilde \chi}^{i}= \pa_{\mu}{\tilde \chi}^{i} + 
g^{{\hat i}k}\pa_{\hat l}g_{k\hat r}
{\tilde \chi}^{r}\pa_{\mu}B^{l}\nonumber \\
R_{{\hat i}m{\hat j}n}=\pa_{\hat i}\pa_{m}g_{n{\hat j}}-
g^{k{\hat l}}\pa_{\hat i}g_{k{\hat j}}\pa_{m}g_{n{\hat l}}\;.
\eq
The superfields $\F^{i}$ and $\X^{i}$ span a 
riemannian manifold ${\cal M}$ 
\footnote{Equivalently the $\q$, $\qwt$ components, 
$A^{i'}s$, $B^{i'}s$, span the same manifold and we 
can use indistinctly $(\F^{i},\X^{i})$ or $(A^{i},B^{i})$
to denote local coordinates on ${\cal M}$.} whose 
metric comes from the kinetic term of 
the scalar fields in
(\ref{action1}), that is  
\be 
g_{IJ} = \left( \begin{array}{cc} \0 & g_{i\hat{\jmath}} \\
g_{\hat{\imath} j}& \0 \
\end{array} \right)\;,\;\;\;
g_{i\hat{\jmath}}=\frac{\pa^2 K}{\pa A^{i}\pa B^{j}}\;.
\label{metrica} 
\ee 
Adopting a complex coordinates system $(Z^{i},Z^{\ov i})
\equiv (A^{i}+iB^{i}, A^{i}-iB^{i})$ for 
the manifold spanned by the scalar superfields 
we get a metric that is 
non-hermitian. In order to characterize the 
geometry which lies under our construction we have 
also to analyse how the action
behaves under a transformation of coordinates. 
Let us consider ``holomorphic'' transformations, i.e., 
\bq
(\F^{i},\X^{i})\rightarrow (\F^{'i},\X^{'i}) 
\equiv(\F^{'i}(\F)=e^{\l.\k}\F^{i},\X^{'i}(\X)= e^{\l.\t}\X^{i}) 
\label{isometria}\;, 
\eq
with ${{\cal K}_{a}}= (\k_{a}(\small{\F)},\t_{a}(\small{\X})) 
\equiv(\k_{a}^{i}\pa_{i}, \t_{a}^{i}\pa_{\hat i})$ 
holomorphic killing vectors. This is equivalent to 
the existence of a locally-product structure on the manifold 
${\cal M}$ \cite{yano},
i.e. the existence of a mapping on the tangent space of ${\cal M}$ 
satisfying, 
\bq 
& I: T{\cal M} \rightarrow T{\cal M} 
\nonumber \\ 
& I^{2}=1\;. 
\label{estrutura} 
\eq 
The holomorphicity of the killing vectors 
is a consequence of imposing the
vanishing of the Lie derivative of $I$ 
along ${\cal K}$: ${\cal L}_{\cal K} I=0$. 
In the canonical coordinate system
defined by $X^{I}$ we have the locally-product 
structure written as 
\be 
I^{I}\,_{J} = \left( \begin{array}{cc} 
I^{i}\,_{j} &\0 \\
0 & I^{\hat {\imath}}\,_{\hat {\jmath}} 
\end{array} \right) =
\left( \begin{array}{cc} 
\d^{i}_{j} &\0 \\ 
\0 & -\d^{\hat i}_{\hat j} 
\end{array}\right)\;\;\;. 
\label{estrutura1} 
\ee 
The metric (\ref{metrica}) is antidiagonal because of the
relation,
\footnote{The fact of an antidiagonal metric forbids the 
manifold of being a locally-product space. In others term: 
in a locally-product space we have 
$I^{I}_{M}I^{J}_{N}g_{IJ} = g_{MN}$ 
instead of (\ref{antidiagonal}).} 
\be
I^{I}_{M}I^{J}_{N}g_{IJ} = -g_{MN}\;. 
\label{antidiagonal} 
\ee 
This allows us to define a sympletic 2-form 
in the same way
as for a K\"ahlerian manifold, 
$w=(w_{IJ})\equiv Ig$. 
Since $w$ is closed, the metric 
is derived by a
scalar function $K(\F,\X)$ according to 
(\ref{metrica}).Therefore, we have 
the following assertion:\\
\noindent 
{\it{The N=1 manifold, i.e., the target manifold 
associated to the N=1 real 
supersymmetric $\s$-model is a 2n-dimensional 
Riemannian manifold that admits a
locally-product structure $I$, a metric $g$ and a 
2-form $w$ such that}}: \\
\noindent 
(i) $g (IU,IV)=-g(U,V)$ 
({\it{the metric is ``anti-hermitian''}}) \\
\noindent 
(ii)$w(U,V)=g(IU,V)$ {\it{is closed}}. 

It should be observed that all the geometric
content of the manifold ${\cal M}$ is encoded 
in the following assumptions:\\
\noindent
(i) we have an action given by (\ref{action})
\\
\noindent
(ii) the coordinates transform holomorphically. \\
\noindent
The manifold ${\cal M}$ obviously shares common properties of a
Kh\"aler and a locally-product manifold.  
Like a Kh\"aler manifold, 
${\cal M}$ also admits a metric that is hybrid \cite{yano} 
and there is also a symplectic 2-form $w$ that fix the form of the
metric as derivatives of a scalar potential $K$. The similarity 
with a locally-product space comes from the existence in
both of them of a locally-product structure. 

The Levi-Civitta 
conexion on ${\cal M}$ assumes the same form as in the
Kh\"alerian case, 
$\G^{H}_{IJ} \equiv (\G^{h}_{ij}, \G^{\hat h}_{{\hat i}{\hat j}})$ 
where, $\G^{h}_{ij}=g^{h{\hat r}}
\pa_{i}g_{j{\hat r}}\,\; \G^{\hat h}_{{\hat i}{\hat j}}= 
g^{{\hat h}r}\pa_{\hat i} g_{{\hat j}r}$. 
The canonical locally-product
structure has zero covariant derivative $\nabla I=0$ 
and it is also integrable \cite{yano}.

\section{The Real N=2 Supersymmetric $\s$-Model}
The N=2 supersymmetric transformation of the model 
can be realized
explicitly in a superspace N=1 if we write 
it as follows \cite{hull},\cite{marcelo2}, 
\cite{gates3},\cite{helayell}: 
\bq
\d\F^{i}= {\wt{D}}^{2}(\e \O^{i})\;,\;\; 
\d\X^{i}=D^{2}(\z \cv^{i})\;, 
\label{susy1} 
\eq
where $\O^{i}=\O^{i}(\F,\X)$ and
$\cv^{i}=\cv^{i}(\F,\X)$ are considered
as generic functions of superspace for a moment but  
later on they will be related to the split
quaternionic structures. The parameters 
$\e$ and $\z$ are real superfields. The 
requirement of being a
supersymmetry transformation implies that 
\bq 
\d_{1}\d_{2} -\d_{2}\d_{1}\approx \partial 
\eq 
and this gives us the
relations 
\bq 
&\O^{i},_{j{\hat k}}\O^{j},_{\hat r} - 
\O^{i},_{j{\hat r}}\O^{j},_{\hat k}=0\;,\;\;\; 
\cv^{i},_{{\hat j}r}\cv^{j},_{k} -
\cv^{i},_{{\hat j}k}\cv^{j},_{r}=0 
\label{integrabilidade}
\\ 
&\O^{i},_{\hat j}\cv^{j},_{k}= \d^{i}_{k}\;,\;\;\; 
\cv^{i},_{j}\O^{j},_{\hat k}=\d^{i}_{k} 
\label{estruturaprodutolocal}
\\
&{\wt{D}}^{2}\O^{i}=0\;,\;\;\;D^{2}\cv^{i}=0 
\\ 
&{\wt{D}}^{2}\e=0\;,\;\;\;D_{\a}\e=0\;,\;\;\;\pa_{\mu}\e=0 
\label{antiquiral}
\\ 
&D^{2}\z=0\;,\;\;\; {\wt{D}}_{\ad}\z=0\;,\;\;\;\pa_{\mu}\z=0\;. 
\label{quiral} 
\eq 
Eqs.(\ref{antiquiral},\ref{quiral}) determine the parameters 
$\e$ and $\l$ as respectively spacetime 
constant antichiral/chiral real superfields. 

The invariance of the action
(\ref{action}) under the N=2 supersymmetry tansformations 
(\ref{susy1}) implies: 
\bq 
&K_{i{\hat j}l}\O^{i},_{\hat m}+
K_{i{\hat j}}\O^{i},_{{\hat m}l}=0 \;,\;\;\; 
K_{{\hat i}j{\hat l}}\cv^{i},_{m}+ K_{{\hat i}j} 
\cv^{i},_{m{\hat l}}=0 
\label{1}
\\ 
&K_{i{\hat j}{\hat l}}\O^{i},_{\hat m}+ 
K_{i{\hat m}}\O^{i},_{{\hat j}{\hat l}}=0 \;,\;\;\; 
K_{{\hat i}jl}\cv^{i},_{m}+ K_{{\hat i}m} 
\cv^{i},_{jl}=0 
\label{2}
\\ 
&K_{i{\hat j}}\O^{i},_{\hat l} + K_{i{\hat l}}\O^{i},_{\hat j}=0\;, \;\;\; 
K_{{\hat i}j}\cv^{i},_{l} + K_{{\hat i}l}\cv^{i},_{j}=0\;.
\label{3} 
\eq 
These set of relations have a geometrical 
interpretation that will be made clear after the 
discussion
on the next section. 
\section{Split-Quaternionic Analysis}
\subsection{Basic Properties of Split Quaternions} 
We will present here some results concerning the 
split-quaternionic algebra. They follow essentially 
the same development of
{\cite{chevalley}}{\cite{kraines}}. Let $H^{'}$ be the 
algebra over $R$ generated by
$[\hat{e}_{0},\hat{e}_{1},\hat{e}_{2},\hat{e}_{3}]$ with 
$\hat{e}_{0}$ being the identity and the others elements 
satisfying
the relations 
\be 
\begin{array}{ccc} 
\hat{e}_{1}\hat{e}_{1}= \hat{e}_{0} 
& \hat{e}_{1}\hat{e}_{2}= \hat{e}_{3}
& \hat{e}_{1}\hat{e}_{3}= \hat{e}_{2} \\ 
\hat{e}_{2}\hat{e}_{1}= -\hat{e}_{3} 
& \hat{e}_{2}\hat{e}_{2}= \hat{e}_{0}
& \hat{e}_{2}\hat{e}_{3}= -\hat{e}_{1} \\ 
\hat{e}_{3}\hat{e}_{1}= -\hat{e}_{2} 
& \hat{e}_{3}\hat{e}_{2}= \hat{e}_{1} 
& \hat{e}_{3}\hat{e}_{3}= -\hat{e}_{0} \;\;. 
\label{alg} 
\end{array} 
\ee 
A generic element of $H^{'}$ is then 
written as 
$\hat{q}=(q^{0},q^{1},q^{2},q^{3})
\equiv q^{0}\hat{e}_{0}+ q^{1}\hat{e}_{1} + 
q^{2}\hat{e}_{2} + q^{3}\hat{e}_{3},\;
q^{0},...,q^{3}\in R$. 
Addition and product of elements in $H^{'}$ are 
given naturally as 
$\hat{q}+ \hat{p}=(q^{0} + p^{0},...,q^{3}+p^{3})$ 
while the product is defined by (\ref{alg}). 
The multiplication by a real $\l$ is given by 
$\l \hat{q}= \hat{q}\l =
(\l q^{0},...,\l q^{3})$. Complex conjugation on 
$H^{'}$ is defined in the following way:
$\hat{q}=(q^{0},q^{1},q^{2},q^{3})
\longrightarrow \hat{q}^{*}=(q^{0},-q^{1},-q^{2},-q^{3})$ 
and it satisfies, 
$(\l \hat{q} + \z \hat{p})^{*} = \l \hat{q}^{*} + 
\z \hat{p}^{*},\; (\hat{p}\hat{q})^{*}= \hat{q}^{*} \hat{p}^{*},\; 
((\hat{q}^{*})^{*})=\hat{q}$. 
In particular, 
$\hat{q}\hat{q}^{*}=\hat{q}^{*}\hat{q}= 
({q^{0}}^{2}-{q^{1}}^{2}-{q^{2}}^{2}+{q^{3}}^{2})\hat{e}_{0}$. 
The norm of
$\hat{q}$ is the real number 
$|\hat{q}|\equiv {q^{0}}^{2}-{q^{1}}^{2}-{q^{2}}^{2}+{q^{3}}^{2},$ 
that can be zero even if
$\hat{q}\ne {\bf 0}$, so 
that $H^{'}$ is not a divison algebra. 

Consider now $H^{'n}\equiv H^{'}\times ...\times H^{'}$ 
as the set of elements ${\cal Q}=
(\hat{q}^{1},..., \hat{q}^{n}),\; \hat{q}^{i} \in H^{'}, i=1,...n.$ 
We endow $H^{'n}$ with a structure of right $H^{'}$-module by
defining the operations ${\cal P} + {\cal Q}= 
(\hat{p}^{1}+\hat{q}^{1},..., \hat{p}^{n}+ \hat{q}^{n}),\;
{\cal Q}\hat{p}=
(\hat{q}^{1}\hat{p},...,\hat{q}^{n}\hat{p}),\; \forall 
{\cal P},\;{\cal Q}\in H^{'n},\; \forall \hat{q}\in H^{'}$. 
From these definitions we have the following properties: 
$({\cal P} + {\cal Q})\hat{q}= {\cal P}\hat{q} + 
{\cal Q}\hat{q}, \; {\cal Q}(\hat{p}+ \hat{q})={\cal Q}\hat{p} + 
{\cal Q}\hat{q},\;{\cal Q}(\hat{p}\hat{q})= ({\cal Q}\hat{p})\hat{q}.$ 
$H^{'n}$ is
generated by the elements 
$\{ E_{1},..., E_{n}\},\; E_{i}=(\delta_{ij}\hat{e}_{0}), j=1,...,n.$ 
Then we have 
${\cal Q}=\sum E_{i}\hat{q}^{i}$. 

The symplectic product on $H^{'}$ is a bilinear map 
$<\;\;,\;\;>:H^{'n}\times H^{'n}\longrightarrow
H^{'},\;\;$ defined by
$({\cal P},{\cal Q})\longrightarrow <{\cal P},{\cal Q}>= 
\sum \hat{p}^{i}\hat{q}^{i*}$, 
and satisfies, 
$<{\cal P}{\hat \l},{\cal Q}>=<{\cal P},{\cal Q}{\hat \l^{*}}>.$ 

An endomorphism in $H^{'n}$ is a mapping 
$\s:H^{'n}\longrightarrow H^{'n}$, such that 
 $\s({\cal P}_{1}+{\cal P}_{2})=\s({\cal P}_{1})+ 
 \s({\cal P}_{2}),\, \s({\cal P}{\hat q})=
 \s({\cal P}){\hat q}$ and, in
particular, a linear endomorphism is completely 
determined when it is given its action on the basis $\{E_{i}\}$,
$\s(E_{i})\equiv E_{j}{\s}_{ji}$, therefore 
$\s{\cal P}=\sum E_{i}{\s}_{ij}{\hat p}^{j}$. 
The association $\s \leftrightarrow \s_{ij}$ is
a bijection and it allows us to represent the 
action of a linear endomorphism in $H^{'n}$ by means of 
the matrices $\s \equiv (\s_{ij})\in {\cal M}_{n\times n}(H')$ 
with coeficients in $H^{'}$.

Every linear endomorphism of 
$H^{'n}$ preserving
the symplectic form $<,>$ is said to be a symplectic 
transformation. The set of such transformations defines the
symplectic group. It is convenient to deal with the 
symplectic product as a bilinear in $C^{2n}$,
since this will permit us to characterize the symplectic 
group in terms of a subgroup of $GL(2n,C)$, the so-called
linear split-symplectic group. 

\subsection{The Linear Split-Symplectic Group} 
In order to define the linear
split-symplectic group we remember that 
${\tilde H}=[{\hat e}_{0},{\hat e}_{3}]$ 
is a subalgebra of $H$ that has inverse.
${\tilde H}$ is also isomorphic to $C$ 
by the map: ${\hat q}=q^{0}{\hat e}_{0} +q^{3}{\hat e}_{3} 
\longleftrightarrow z = q^{0} +iq^{3}$. 
Then, we can define the action of $C$ on $H^{'}$ 
as $({\hat q},z)\longrightarrow {\hat q} z:=
(q^{0}x-q^{3}y){\hat e}_{0} + (q^{1}x - q^{2}y){\hat e}_{1} +
(q^{1}y+q^{2}x){\hat e}_{2} + (q^{0}y + q^{3}x){\hat e}_{3},\;
(z=x+iy).$ 
We can also write ${\hat q}_{k}\equiv {\hat e}_{0} z_{k}^{*} - 
{\hat e}_{1}z_{k+n}$, with $z_{k}= q^{0}_{k}{\hat e}_{0}
-q^{3}_{k}{\hat e}_{3},\; z_{k+n}=-q^{1}_{k}{\hat e}_{0} - 
q^{2}_{k}{\hat e}_{3},$ that associates ${\cal Q}=({\hat q}_{i})\in H^{'n}$ 
with
${\tilde {\cal Q}}=Z=(z_{\hat k})\equiv (z_{k},z_{k+n})\in C^{2n}.$ 
Let us consider then 
${\cal Q}=({\hat q}_{i}),\; {\cal P}=({\hat p}_{i})\in H^{'n}$, 
to which corresponds 
${\tilde {\cal Q}}=(z_{\hat k})\equiv (z_{k},z_{k+n}),\;
{\tilde {\cal P}}=(w_{\hat k})\equiv (w_{k},w_{k+n})\in C^{2n}.$ 
Given the symplectic transformation $\s=(\s_{ij})$ we have associated
a transformation ${\tilde \s}$ of $GL(2n,C)$ in $C^{2n}, 
{\tilde {\cal Q}}\longrightarrow {\tilde \s} {\tilde {\cal Q}}$.
Since $\s$ is symplectic we have that $<\s{\cal Q},\s{\cal P}>=
<{\cal Q},{\cal P}> \Longrightarrow$ 
\bq 
&{\hat e}_{0} \sum_{{\hat k},{\hat r},{\hat l},
{\hat s}=1}^{2n}
z^{*}_{\hat r} {\tilde \s}^{\dagger}_{{\hat r}{\hat k}} 
I_{{\hat k}{\hat l}}{\tilde \s}_{{\hat l}{\hat s}} w_{{\hat s}} +
{\hat e}_{1}\sum_{{\hat k},{\hat r},{\hat l},{\hat s}=1}^{2n} z_{\hat r} 
{\tilde \s}^{t}_{{\hat r}{\hat k}} J_{{\hat k}{\hat l}}{\tilde \s}_{{\hat l}
{\hat s}} w_{{\hat s}}= {\hat e}_{0} 
{\sum_{{\hat k},{\hat l}=1}^{2n}} z^{*}_{\hat k} I_{{\hat k}{\hat l}} 
w_{{\hat l}} + 
{\hat e}_{1}{\sum_{{\hat k},{\hat l}=1}^{2n}} z_{\hat k} 
J_{{\hat k}{\hat l}} w_{{\hat l}} \nonumber
\Longrightarrow \\
&\Longrightarrow \left \{ {\begin{array}{ccc} 
&{\tilde \s}^{\dagger} I {\tilde \s} = I \\ 
&{\tilde \s}^{t}J {\tilde \s} =J
\end{array}} \right. 
\eq 
where 
\be 
I=(I_{{\hat k}{\hat l}}) = 
\left( \begin{array}{cc} 
\d_{kl} &\0 \\ \0 & -\d_{kl}
\end{array} \right), \;\;\;
J=(J_{{\hat k}{\hat l}}) = 
\left( \begin{array}{cc} 
\0 & \d_{kl} \\ -\d_{kl} & \0
\end{array}\right)\;\;\;. 
\label{estrutura} 
\ee 
In particular one has, 
${\cal P'}={\cal P}{\hat \l}\rightarrow 
{\tilde {\cal P'}}=(\l^{0}{\bf 1}+ \l^{i} (-1)^{i+i} {\cal I}_{i}) 
{\tilde {\cal P}}$ where 
\be ({\cal I}_{i})=\left( \left( \begin{array}{cc} 
\0 & -{\bf 1} \\ 
-{\bf 1} & \0 \end{array}\right),\; 
\left( \begin{array}{cc} 
\0 & -{\bf i} \\ {\bf i} & \0 
\end{array}\right),\;
\left( \begin{array}{cc} 
-{\bf i}& \0 \\ \0 &{\bf i} 
\end{array}\right)\; \right) 
\ee 
represent the split quaternionic algebra in $C^{2n}$. 
\subsection{The Fundamental 4-Form}
Let us define now 2-forms on $H^{'n}$ by writing 
\bq 
&\o_{i}({\cal P},{\cal Q}) 
\equiv \frac{1}{2} (<{\cal P}{\hat e}_{i},{\cal Q}> + 
<{\cal Q},{\cal P}{\hat e}_{i}>)=\nonumber \\ 
& = \frac{1}{2}\{ {\hat e}_{0} \sum_{{\hat k},{\hat l}=1}^{2n} 
(z^{*}_{\hat k} (-1)^{i+1} ({\cal I}^{\dagger}_{i}I)_{{\hat k}
{\hat l}}w_{\hat l} -w^{*}_{\hat k} (-1)^{i+1} 
({\cal I}^{\dagger}_{i}I)_{{\hat k}{\hat l}} z_{\hat l}) \nonumber \\ 
&+ {\hat e}_{1}
\sum_{{\hat k},{\hat l}=1}^{2n} (z_{\hat k} (-1)^{i+1} 
({\cal I}^{t}_{i}J)_{{\hat k}{\hat l}}w_{\hat l} 
-w_{\hat k} (-1)^{i+1}
({\cal I}^{t}_{i}J)_{{\hat k}{\hat l}} z_{\hat l})\}.
 \eq 
 They have the properties, 
 \bq 
 &\o_{1}({\cal P},{\cal Q})=-\o_{1} 
 ({\cal P}{\hat e}_{1},{\cal Q}{\hat e}_{1})= 
 \o_{1}({\cal P}{\hat e}_{2},{\cal Q}{\hat e}_{2})= 
 -\o_{1}({\cal P}{\hat e}_{3},{\cal Q}{\hat e}_{3}) \\ 
 &\o_{2}({\cal P},{\cal Q})=\o_{2} 
 ({\cal P}{\hat e}_{1},{\cal Q}{\hat e}_{1})= 
 -\o_{2}({\cal P}{\hat e}_{2},
 {\cal Q}{\hat e}_{2})= -\o_{2}({\cal P}{\hat e}_{3},{\cal Q}{\hat e}_{3})\\ 
 &\o_{3}({\cal P},{\cal Q})=\o_{3} 
 ({\cal P}{\hat e}_{1},{\cal Q}{\hat e}_{1})= 
 \o_{3}({\cal P}{\hat e}_{2},{\cal Q}{\hat e}_{2})= 
 \o_{3}({\cal P}{\hat e}_{3},{\cal Q}{\hat
e}_{3})\;\;. 
\eq 
The group $sp(1)$ corresponds to the set 
of unit-split-quaternions, i.e., 
$\{{\hat \l}\in H'; |{\hat \l}|=1\}$ 
and it acts on $\o_{i}$ as 
${\hat \l}\o_{i}\equiv\o_{i} ({\cal P}{\hat \l},{\cal Q}{\hat \l})$. 
It follows then, 
\bq
&\o_{1}({\cal P}{\hat \l},{\cal Q}{\hat \l})= 
({\l^{0}}^{2}-{\l^{1}}^{2}+{\l^{2}}^{2}-{\l^{3}}^{2}) 
\o_{1}({\cal P}{\cal Q})
-2(\l^{0}\l^{3}+ \l^{1}\l^{2})\o_{2}({\cal P},{\cal Q})+ 
\nonumber \\ 
&-2(\l^{0}\l^{2}+ \l^{1}\l^{3})\o_{3}({\cal P},{\cal Q})
\label{l1} 
\eq 
\bq 
&\o_{2}({\cal P}{\hat \l},{\cal Q}{\hat \l})=2(\l^{0}\l^{3}- 
\l^{1}\l^{2})\o_{1}({\cal P},{\cal Q})+
({\l^{0}}^{2}+{\l^{1}}^{2}-{\l^{2}}^{2}-{\l^{3}}^{2})
\o_{2}({\cal P},{\cal Q})+ \nonumber \\ 
&+2(\l^{0}\l^{1}- \l^{2}\l^{3})\o_{3}({\cal P},{\cal Q}) \label{l2} 
\eq 
\bq 
&\o_{3}({\cal P}{\hat \l},{\cal Q}{\hat \l})=
2(-\l^{0}\l^{2}+\l^{1}\l^{3})\o_{1}({\cal P},{\cal Q})+ 
2(\l^{0}\l^{1}+ \l^{2}\l^{3})\o_{2}({\cal P},{\cal Q})+ \nonumber \\
 &+({\l^{0}}^{2}+{\l^{1}}^{2}+{\l^{2}}^{2}+{\l^{3}}^{2}) 
 \o_{3}({\cal P},{\cal Q}) 
 \label{l3} 
 \eq 
 The action of $\s\in sp(n)$ on
$\o_{i}$ is defined by $\o_{i}\rightarrow \s\o_{i}: 
\s\o_{i}({\cal P},{\cal Q})\equiv \o_{i} (\s{\cal P},\s{\cal Q})$ 
and since
the symplectic product is invariant by the action of
 $sp(n)$ we have the forms $\o_{i}$ also invariant by $sp(n)$. 
 We can also define an action of $sp(n)sp(1)$ on $\o_{i}$ as 
 $(\s,{\hat \l})\o_{i}({\cal P},{\cal Q})= \o_{i}(\s{\cal P}
 {\hat \l},\s{\cal Q}{\hat \l})$, where on the right hand 
 side we are supposed to do first the multiplication by $sp(1)$ and
latter by $sp(n)$. 

Finally, we define in $H^{'n}$ a 
4-form $\Ld\equiv \o_{1}\wedge \o_{1} + \o_{2}\wedge \o_{2}
-\o_{3}\wedge \o_{3}$. The action of $sp(n)sp(1)$ is 
defined by the corresponding action of $sp(n)sp(1)$ on each
$\o_{i},\; (\s,{\hat \l})\Ld\equiv (\s,{\hat \l})\o_{1}\wedge 
(\s,{\hat \l})\o_{1} +(\s,{\hat \l})\o_{2}\wedge(\s,{\hat \l})\o_{2}
-(\s,{\hat \l})\o_{3}\wedge(\s,{\hat \l})\o_{3}$, and from eqs. 
(\ref{l1},\ref{l2},\ref{l3}) we have that $\Ld$ is 
invariant
by $sp(n)sp(1)$. 

{\subsection{The N=2 Manifold}}
We follow here the same definitions as was given 
in \cite{ishiara}, but we adapt it to the
split-quaternionic case. Let ${\cal M}$ be a 
smooth 4n-dimensional manifold $(n \ge 1)$ and 
$T{\cal M}$ its tangent
bundle. Consider ${\cal G}$ a 3-dimensional 
subbundle of $Hom(T{\cal M},T{\cal M})$ that has fiber 
${\cal G}_{x}$
 generated by the automorphisms $\{I_{1},I_{2},I_{3}\}$ 
 satisfying the split-quaternionic algebra. The bundle ${\cal G}$ is
called an almost split-quaternion structure in ${\cal M}$ and 
$({\cal M},{\cal G})$ is an almost split-quaternion
manifold. 

${\cal M}$ 
admits a metric $g$ such that
$g(sV,V')+g(V,sV')=0$ for all cross-section 
$s\in \G({\cal G})$ and any vector fields 
$V, V'\in T{\cal M}$. This means
$I_{1}$ and $I_{2}$ are almost ``anti-hermitian'' 
relative to $g$ while $I_{3}$ is almost hermitian. 
We call $({\cal M},g)$ an almost split-quaternion metric structure 
and $({\cal M},g,{\cal G})$ an almost
split-quaternion metric manifold. The almost split-quaternion 
structure ${\cal G}$ is said to be integrable if, given any
neighborhood ${\cal U}$ in ${\cal M}$, there exists a system of 
local coordinates 
$X=(x^{k},x^{\hat k})$ in which the split-quaternionic 
structures are written as {\footnote{In \cite{yanomitsue} 
is discussed the conditions for integrability of
split-quaternionic structures. They found that a necessary and 
sufficient condition of integrability is that at least two
of the Niejenhuis tensors $N(I_{k},I_{k})$ and the curvature 
$R$ of the affine connection vanish.}} 
\be 
I_{1}=\left(
 \begin{array}{cc} 
 {\bf 1} & {\bf 0} \\ {\bf 0} & -{\bf 1} 
 \end{array}\right),\;\; 
 I_{2}=
 \left( \begin{array}{cc} 
 \0 & -{\bf 1}
\\ -{\bf 1} & \0 
\end{array}\right),\;\; I_{3}
\left( \begin{array}{cc} 
\0& -{\bf 1} \\ {\bf 1} & \0 
\end{array}\right)\;. 
\ee
An almost split-quaternion manifold which 
have integrable ${\cal G}$ is said to be a 
split-quaternion manifold. 
On ${\cal M}$
 we also define the 2-forms $w_{i}(V,V')= g(I_{i}V,V')$ and the 
 4-form $\Ld=w_{1}\wedge w_{1}+w_{2}\wedge w_{2}
-w_{3}\wedge w_{3}$. 

The analogue definition of a quaternion 
K\"ahlerian manifold also exist in the split quaternionic
case. It is obtained naturally if we impose the condition
$\nabla_{V} s \in \G({\cal G}),\; \forall s\in \G({\cal G})$ and
$\forall V\in T{\cal M}$ with $\nabla$ the Riemannian conection on $M$.  
 This is equivalent to the equations:
 \bq 
 &\nabla_{V} I_{1}= r_{3}(V)
I_{2} + r_{2}(V) I_{3} \nonumber \\ 
&\nabla_{V} I_{2}= -r_{3}(V) I_{1} - r_{1}(V) I_{3} \nonumber \\ 
&\nabla_{V} I_{3}= r_{2}(V)
I_{1} - r_{1}(V) I_{2} \,, 
\eq 
with $r_{i}$ being 1-forms. This set of equations are 
also equivalent to the condition
$\nabla_{V}\Ld=0$. 

We define now the N=2 manifold, 
i.e., the target manifold of the N=2 real $\s$-model. It
constitutes the extension of hyperK\"ahler manifold to 
to the split-quaternionic 
case. Let $({\cal M},g,{\cal G})$ be a
split-quaternion metric manifold. It will define an N=2 manifold iff, 
$\forall x\in {\cal M},\; {\cal G}_{x}$ satisfies 
$\nabla_{V} I_{i}=0\;, i=1,2,3$. 

Now, it is straightforward to show that this 
is indeed the manifold satisfying the N=2
supersymmetry constraints (\ref{integrabilidade}-\ref{3}). 
Indeed, from the superfields 
$\O,\;\cv$ introduced in (\ref{susy1}) we identify 
\be
I_{2}=\left( \begin{array}{cc} {\bf 0} & \O^{i}_{,\hat j} \\ 
\cv^{i}_{,j} & {\bf 0} \end{array}\right)\;, 
\ee 
$I_{1}$ is given by (\ref{estrutura1}) and
$I_{3}\equiv I_{1}I_{2}$. Eq.(\ref{integrabilidade}) is associated 
to the integrability condition and corresponds 
to the requirement of
$N(I_{2},I_{2})=0$. Eq.(\ref{estruturaprodutolocal}) 
comes from the fact 
that ${\cal G}$ is a split-quaternion structure and so that 
$I_{2}^{2}=1$.
Eqs.(\ref{1},\ref{2}) corresponds to $\nabla_{V}I_{2}=0$ 
(with the Riemannian connection restricted to
 the Levi-Civitta) and finally 
eq.(\ref{3}) is a consequence of $g(sV,V')+g(V,sV')=0$, i.e. of $I_{2}$
 be anti-hermitian relative to $g$. 
 
 \section{Concluding Remarks}
The geometric content of the real models obtained 
here presents new features such as
a locally-product structure instead of a complex 
structure in the N=1-model and split-quaternionic structures
replacing the quaternionic ones in the N=2-extension. 
The emergence of this geometric structure is determined only
by the physical requirements that there is an action 
which is supersymmetric invariant and that the transformations of the 
scalar superfields being restricted to be holomorphic.
In these real models, the possible couplings 
with vector superfields no longer correspond to simply
gauging the isometry group since now the full set of
locally-product structure 
does not leave the metric invariant. It may be even
possible that further restrictions can arise in order to achieve a 
gauge-invariant model. Also, the characterisation of
those manifolds in terms of holonomy groups (see \cite {salamon}) 
can also be developed and compared with the
definitions we have gotten from a purely tensorial analysis, the 
starting point to this being the construction of the 
fundamental 4-form $\Ld$. 
Finally, the analysis of a similar process of generating new
hyperK\"ahler manifolds using the quotient process of \cite{hitchin} 
would deserve some investigation.
\section*{Acknowledgements} 
M.C. thanks Profs.
T.Kori, K.Ozawa, and T.Suzuki and Y.Homma from the mathematics
department of Waseda University 
for many discussions. MC is also grateful to 
the japanese monbusho for his
schoolarship and to Mr. Hassu and all the staff of 
the Centro Cultural e Informativo of the japanese Consulate at Rio
de Janeiro. Finally, he expresses his gratitude to 
Miss Ying Chen and Aruvaru Buraun for many helpful conversations.
 
\end{document}